\documentclass[fleqn,usenatbib]{mnras}

\usepackage{newtxtext,newtxmath}
\usepackage[T1]{fontenc}

\DeclareRobustCommand{\VAN}[3]{#2}
\let\VANthebibliography\thebibliography
\def\thebibliography{\DeclareRobustCommand{\VAN}[3]{##3}\VANthebibliography}

\usepackage{graphicx}    % Including figure files
\usepackage{amsmath}    % Advanced maths commands
\newcommand{\hhh}{$^{\mathrm{h}}$}
\newcommand{\mmm}{$^{\mathrm{m}}$}

\newcommand{\ddd}{$^{\mathrm{\circ}}$}

\title[MeerKAT imaging of 47 Tucanae]{A new pulsar candidate in 47 Tucanae discovered with MeerKAT imaging}

\author[I. Heywood]{
Ian Heywood$^{1,2,3}$\thanks{E-mail: ian.heywood@physics.ox.ac.uk}
\\
$^{1}$Astrophysics, Department of Physics, University of Oxford, Keble Road, Oxford, OX1 3RH, UK\\ 
$^{2}$Centre for Radio Astronomy Techniques and Technologies, Department of Physics and Electronics, Rhodes University, PO Box 94, Makhanda 6140, South Africa\\
$^{3}$South African Radio Astronomy Observatory, 2 Fir Street, Black River Park, Observatory 7925, South Africa\\
}

\date{Accepted 2023 July 04. Received 2023 July 04; in original form 2023 June 28}

\pubyear{2023}

\begin{document}
\label{firstpage}
\pagerange{\pageref{firstpage}--\pageref{lastpage}}
\maketitle

% Abstract of the paper
\begin{abstract} % 200 words

\noindent
MeerKAT imaging of the globular cluster 47 Tucanae (47 Tuc) reveals 1.28~GHz continuum emission at the locations of 20 known millisecond pulsars (MSPs). We use time series and spectral imaging to investigate the image-domain characteristics of the MSPs, and search for previously unknown sources of interest. The MSPs exhibit a range of differences in their temporal and spectral properties compared the general background radio source population. Temporal variability differs strongly from pulsar to pulsar, some appearing to vary randomly on 15 min timescales, others varying coherently by factors of $>$10 on timescales of hours. The error in the typical power law fit to the spectrum emerges as a powerful parameter for indentifying the MSPs. This behaviour is likely due to differing diffractive scintillation conditions along the sight lines to the MSPs. One MSP exhibits tentative periodic variations that are consistent with modulation due the orbit of an eclipsing binary system. One radio source has spectro-temporal properites closely resembling those of the MSP population in the cluster, and we report its position as a candidate new MSP, or alternatively an interferometric localisation of one of six MSPs which do not yet have an accurate position from the timing solutions.

\end{abstract}

% Select between one and six entries from the list of approved keywords.
% Don't make up new ones.
\begin{keywords}
globular clusters: individual: 47 Tucanae -- pulsars: general -- radio continuum: general
\end{keywords}

%%%%%%%%%%%%%%%%%%%%%%%%%%%%%%%%%%%%%%%%%%%%%%%%%%

%%%%%%%%%%%%%%%%% BODY OF PAPER %%%%%%%%%%%%%%%%%%

\section{Introduction}

Globular clusters (GCs) contain high numbers of millisecond pulsars (MSPs) which are spun-up via the high interaction rates with other stars in these extremely dense environments \citep{ransom2008}. The recent deployment of next-generation radio interferometers such as the sensitive and wide-field Square Kilometre Array precursor instruments ASKAP \citep{hotan2021} and MeerKAT \citep{jonas2016} has led to renewed interest in the prospects of using image-domain data to discover new pulsars, and to expand the parameter space for the discovery of variables and transienst in general \citep[e.g.][]{fender2016,murphy2021}. Indeed, the first millisecond pulsar was discovered following an investigation of an unusual feature in a radio continuum image which was arising due to interplanetary scitillation \citep{backer1982}, and at the other end of the periodicity scale, recent results have demonstrated the potential for using image-domain observations to detect potential neutron star systems that are emitting radio pulses with periodicities that are beyond the death lines in the period--period-derivative ($P$--$\dot{P}$) parameter space \citep{caleb2022,hurleywalker2022}.

Motivated by this, we present an analysis of archival MeerKAT imaging observations of 47 Tucanae (47 Tuc), and present the first MeerKAT image of this GC. 47 Tuc is located at a distance of 4.52~$\pm$~0.03~kpc \citep{baumgardt2021} and contains 29 known MSPs\footnote{\url{https://www3.mpifr-bonn.mpg.de/staff/pfreire/GCpsr.html}}, of which 23 have published positions \citep[e.g.][]{manchester1991,robinson1995,camilo2000,freire2001,pan2016,ridolfi2016,freire2017,abbate2023}. Due to this high number of known MSPs we can investigate their characteristics in the image domain data, in both the time and frequency integrated `continuum' image, as well as by generating sub-images of the data along those dimensions. From there we can attempt to search for previously uncharacterised sources that share these properties, as well as search for other sources of interest that exhibit unusual properties when compared to the general (predominantly extragalactic) compact source population. Note that throughout this letter we refer to the pulsars in 47 Tuc using their alphabetical designation, and omit the J0023$-$7204 or 47 Tuc prefix.

\section{Observations and data products}
\label{sec:obs}

We make use of publicly available data from the MeerKAT radio telescope, taken during its science verification (SV) phase, see Table \ref{tab:meerkat_obs} for details. The calibration strategy during SV was conservative, with frequent scans of calibrator sources. The secondary calibrator was observed for 1.73 min for every 14.8 min target scan, and the primary calibrator was visited approximately once per hour, and observed for 9.7 min. The data were recorded in full polarisation mode, and scans of a primary polarisation calibrator were included, however we did not perform full-Stokes imaging for this study.

\begin{table}
%\begin{minipage}{175mm}
\centering
\caption{Summary of the MeerKAT observations 47 Tuc.}
\begin{tabular}{ll} \hline 
Observation date         & 2018-06-17         \\ 
Block ID                 & 1529209968         \\
Proposal ID              & SSV-20180615-FC-01 \\
Block duration           & 9.14~h             \\ 
On-source time           & 5.17~h             \\ 
Frequency range          & 0.856 -- 1.712 GHz \\
Number of channels       & 4096               \\
Number of antennas       & 62                 \\
Integration time per visibility & 8~s         \\
Correlation products     & 4 (HH, HV, VH, VV) \\
Target position (J2000)  & 00\hhh24\mmm05\fs36 $-$72\ddd04\mmm53\farcs2 \\
Primary calibrator       & PKS 0408$-$65      \\
Secondary calibrator     & J0252$-$7104       \\ 
Polarization calibrator  & 3C138              \\\hline
\end{tabular}
\label{tab:meerkat_obs}
%\end{minipage}
\end{table}

The data were retrived from the archive in Measurement Set (MS) format using the KAT Data Access Library\footnote{{\sc katdal}; \url{https://github.com/ska-sa/katdal}}, and averaged by a factor of 4 in frequency, resulting in 1024 channels. The MS was then processed using {\sc oxkat}\footnote{v0.3; \url{https://github.com/IanHeywood/oxkat}} \citep{heywood2020}, a set of scripts for automatically processing MeerKAT continuum data. The process involves cross-calibration using {\sc casa} \citep{casa2022}, automatic removal of radio frequency interference using {\sc casa} and {\sc tricolour} \citep{hugo2022}, self-calibration using {\sc cubical} \citep{kenyon2018}, direction dependent calibration using {\sc killms} \citep{smirnov2015}, and imaging and deconvolution performed using either {\sc wsclean} \citep{offringa2014} or {\sc ddfacet} \citep{tasse2018}. Primary beam correction is applied using an azimuthally-averaged model of the Stokes I beam produced using the {\sc katbeam} package\footnote{\url{https://github.com/ska-sa/katbeam}}. Diagnostic plots are provided by {\sc ragavi}\footnote{\url{https://github.com/ratt-ru/ragavi}} and {\sc shadems} \citep{smirnov2022}, and all packages are containerised using {\sc singularity} \citep{kurtzer2017}. Full details and validation of the data processing workflow are provided by \citet{heywood2022}. 

\begin{figure}
\centering
\includegraphics[width= \columnwidth]{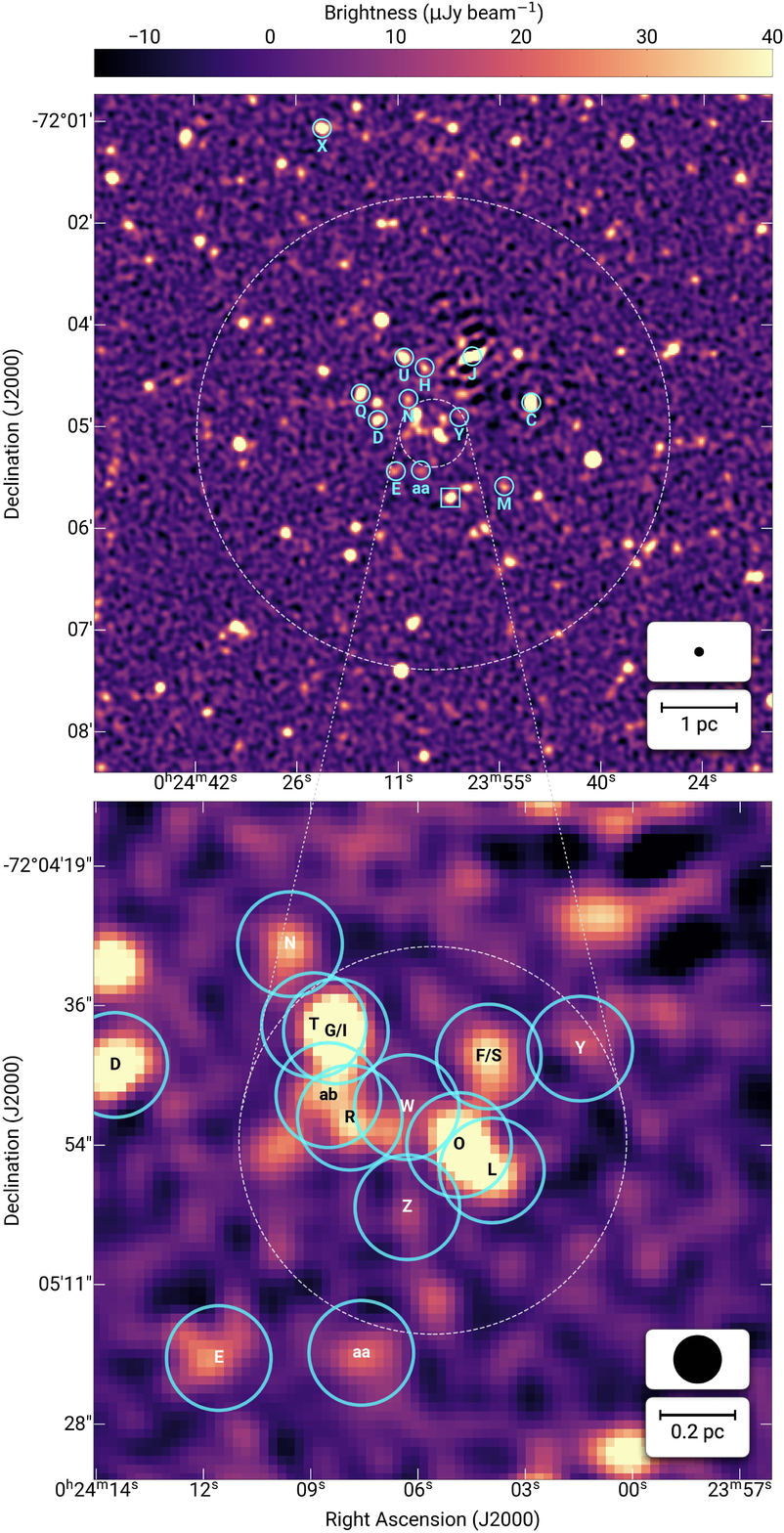}
\caption{The MeerKAT image of 47 Tuc at 1.28~GHz. The inner dashed circle on the upper panel shows the cluster core \citep[radius 24$''$;][]{howell2000} with the outer dashed circle corresponding to the half mass radius \citep[167\farcs4;][]{harris1996}. The cyan markers and labels mark the positions of 23 MSPs with published coordinates, 20 of which are coincident with significant features in the continuum image. The lower panel shows a zoom of the core region showing the high concentration of MSPs within. Note the five pulsars that appear as fully or partially blended components at this angular resolution. The circular marker in the lower right of each panel shows the circular Gaussian restoring beam, which has a FWHM of 5\farcs6, and the accompanying scale bar shows the coversion of angular extents to physical sizes at the distance of 47 Tuc. The square marks the parameters of the candidate MSP presented in Section \ref{sec:newpulsar}.}
\label{fig:maps}
\end{figure}

A time-integrated, multi-frequency synthesis (MFS) image of 47 Tuc is presented in Figure \ref{fig:maps}, with details of the annotations provided in the figure caption. Continuum emission is seen at the position of 20 of the 23 pulsars in 47 Tuc, with non-detections of W, Y, and Z, although the latter two have hints of associated emission. Additionally, at this angular resolution we are unable to separate the emission for the groupings G/I and also F/S, with T also partially overlapping with G/I. The components associated with ab/R and O/L are also partially blended. Thirteen MSPs have clear, isolated radio continuum detections.

Figure \ref{fig:maps} shows an intriguing feature, in that the pulsars C and J have not deconvolved properly. Calibration deficiencies, in particular those due to direction dependent effects \citep[DDEs; see e.g.][]{smirnov2011a,smirnov2011b}, often result in residual PSF-like structures around the brighter sources. The dominant DDE for MeerKAT L-band obsevations is related to the antenna primary beam pattern coupled with pointing errors. These effects do not have a strong manifestation close to the pointing centre, and sources of comparable brightness to MSPs C and J do not suffer similar deficiencies. One could thus make the prediction at this stage that these sources are strongly time-variable over the course of the observation. Intrinsic variability of a particular radio source during the observation is analogous to a direction-dependent amplitude error in the eyes of the calibration and imaging process. The presence of persitent, unexpected PSF-like patterns around a source is a good indicator that its radio emission may be variable, either intrinsically or due to propagation effects \citep[e.g.][]{oosterloo2020}.

We restrict the analyses that follow to the inner 0.625~$\times$~0.625 deg$^{2}$ of the image, covering up to 7 times the half mass radius from the cluster centre. The primary beam gain in this region is no less than 0.75. The main lobe of the MeerKAT Stokes I beam is not azimuthally symmetric, and our selection limits any instrumental source variability induced by rotation of the beam pattern with respect to the sky to $<$1 per cent.

To extract a `control' sample of sources -- a representative sample of the general source population, predominantly background extra galactic sources, but that may also include new pulsar candidates -- we used the {\sc pybdsf} source finder \citep{mohan2015} with its default settings. In the absence of confusion or significant nebular emission, a pulsar will be detected as true point source. The control sample was therefore filtered to exclude extended sources by enforcing a peak brightness to integrated flux density ratio 0.9~$<$~S$_{\mathrm{peak}}$~/~S$_{\mathrm{int}}$~$<$~1.0, and selecting only single point / Gaussian components with S$_{\mathrm{peak}}$~$>$~ 10~$\mathrm{\mu}$Jy beam$^{-1}$. Following this the control sample contains 927 sources. Detections of known pulsars were flagged, and any known pulsar that was not picked up by {\sc pybdsf} was fitted using the {\sc casa} {\sc imfit} task. 

In order to study the temporal properties of the pulsars and other sources in the field in the image domain, we used {\sc ddfacet} to produce deconvolved, full-band MFS images with direction-dependent corrections applied for each of the 21 $\times$ 14.8 min scans. We extract light curves from this sequence of images for both MSPs and control sources by force-fitting a Gaussian with the shape of the epoch-specific restoring beam at the position of each source using the {\sc casa} {\sc imfit} task. 

Spectral properties of the sources are examined by means of a spectral index\footnote{We adopt that convention that the power law $S~\propto~\nu^{\alpha}$ relates the spectral index $\alpha$ to the flux density (or peak brightness) $S$.} map and an accompanying error map. These are made by imaging and deconvolving the data in 8 sub-bands. The angular resolutions of the resulting stack of images are homogenised by convolving the images to a marginally coarser resolution than that of the lowest sub-band (8\farcs5). We then apply primary beam correction appropriate for each frequency. Following this, we fit for the linear gradient ($\alpha$) of each sight-line through the resulting cube in log-$\nu$ / log-$S$ space. These values are then recorded as a spectral index and spectral index error ($\sigma_{\alpha}$) map. For each source we weight the spectral index and spectral index error map with a circular 8\farcs5 Gaussian centred on its position, and the brightness-weighted mean values of $\alpha$ and $\sigma_{\alpha}$ are then associated with each component in the catalogue for further analysis.

% \subsubsection{Probability of chance alignment with an extragalactic radio source}

% The faintest MSP counterpart has a peak brightness of X. The surface density of extragalactic radio sources at this brightness or greater is Y (Wilman). The probability of an extragalactic continuum source within one synthesised beam width is therefore Z.

\section{Discussion}

\begin{figure*}
\centering
\includegraphics[width=1.0 \textwidth]{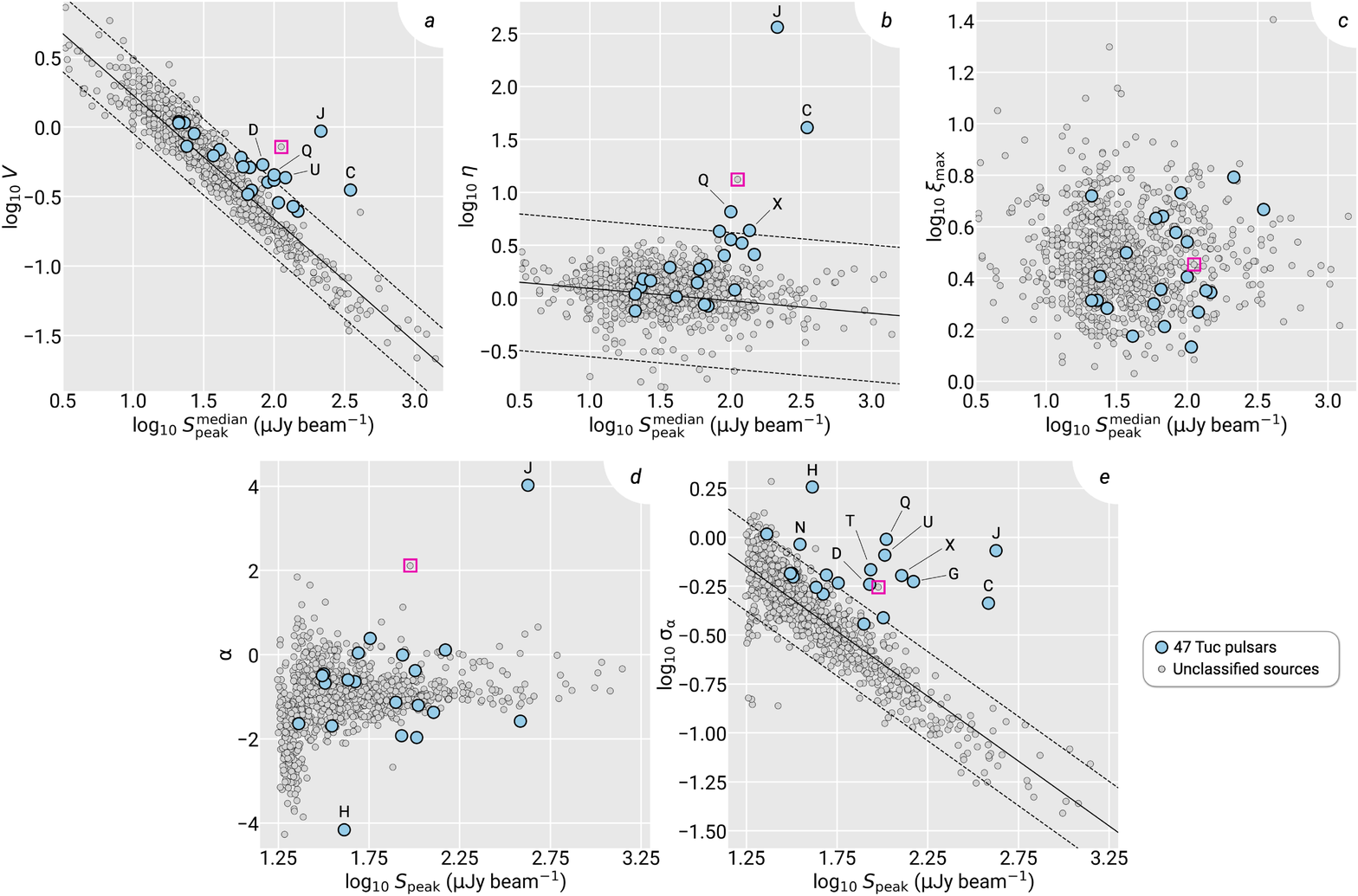}
\caption{Panels $a$, $b$ and $c$ show the variability statistics $V$, $\eta$ and $\xi_{\mathrm{max}}$ as a function of the median peak brightness of the light curve of each source (see Section \ref{sec:stats}). Panels $d$ and $e$ show the spectral index and associated standard deviation as a function of the peak brightness in the MFS image (see Section \ref{sec:alpha}). Outlier regions for $V$, $\eta$ and $\sigma_{\alpha}$ are delineated by the lines, which show fits to the distribution and the $\pm$2$\overline{\sigma}$ range, please refer to Section \ref{sec:alpha} for more details. Known MSPs are marked with cyan circles with outliers labeled. The grey dots show the control sample. The pink square highlights the candidate pulsar (see Section \ref{sec:newpulsar}.}
%\vspace{5mm}
\label{fig:stats}
\end{figure*}

\subsection{Temporal variability statistics}
\label{sec:stats}

The light curve for each object can be boiled down to three statistics, which when plotted as an ensemble can reveal interesting behaviour via the identification of outliers \citep[e.g.][]{driessen2020,driessen2022,rowlinson2022,andersson2023}. For a light curve consisting of $N$ brightness measurements, $S_{i}$, the first statistic we make use of is $V$, which is simply the standard deviation of the light curve divided by the mean:
\begin{equation}
V~=~\sigma_{S}~/~\overline{S}.
\label{equ:V}
\end{equation}
The $V$ statistic thus provides some measure of how spread out the values of $S_{i}$ are. The second statistic $\eta$ is based on the reduced $\chi^{2}$ statistic:
\begin{equation}
\eta~=~\frac{N}{N-1}~\sum_{i=1}^{N}~\frac{(S_{i} - \overline{S})^{2}}{\sigma^{2}_{i}}
\label{equ:eta}
\end{equation}
where $\sigma_{i}$ is the standard deviation noise measurement of $S_{i}$. A constant source should therefore have a value of $\eta$ close to unity. The third statistic $\xi_{\mathrm{max}}$ is based on the median absolute deviation (MAD). For each element in the light curve, the residual of the data and the median ($\tilde{S}$) is expressed as a fraction of the MAD:
\begin{equation}
\xi_{i}~=~\frac{S_{i} - \tilde{S}}{\mathrm{median}{\left(|S_{i}-\tilde{S}|\right)}}
\label{equ:xi}
\end{equation}
and $\xi_{\mathrm{max}}$ is the maximum value from the resulting $N$-length set. The $\xi_{\mathrm{max}}$ statistic thus has a higher value for light curves that contain short, burst-like features.

These three statistics are computed for the objects in the control sample, as well as for the known pulsars. The resulting values of $V$, $\eta$, and $\xi_{\mathrm{max}}$ are plotted against the median peak brightness for each object in log-log space on panels $a$, $b$ and $c$ of Figure \ref{fig:stats}. The control sample is plotted in grey, with known MSPs plotted as blue circles. 

Note that the parent catalogues for these figures are drawn from the full time-integrated MFS images described in Section \ref{sec:obs}. Assuming constant sensitivity of the telescope during a given observation, the per-scan snapshot images used to derive the light curves will have noise levels that are 4.6 times higher than the fully integrated image. We will therefore be performing photometric extraction via the fitting of epoch-specific Gaussians at the positions of some sources that are dominated by the thermal noise in the time series imaging. However, this is not an issue in terms of confusing plots of the the variability statistics introduced above, as they are all formed relative to a mean or median baseline. Indeed the correlation between the signal to noise ratio (SNR) of the source brightness and the $V$ parameter is evident in panel $a$ of Figure \ref{fig:stats} (and self-evident in Equation \ref{equ:V}). Similarly, some of the pulsars with fainter continuum detections in the integrated images may also be dominated by noise in the time series imaging. Again we apply force-fitting to extract their light curves to search for brief brightness changes that may not be evident in the time and frequency integrated images.

\subsection{Spectral index properties}
\label{sec:alpha}

As can be be seen in panels $d$ and $e$ of Figure \ref{fig:stats}, spectral index measurements exhibit increasing amounts of scatter as the SNR of the source decreases. Mean $\alpha$ measurements for ensembles of objects as a function of their mean brightness eventually become entirely unreliable due to the imposition of strong selection effects that are coupled to the observational frequency coverage and the intrinsic spectral indices of the source population \citep[see][for a discussion of this]{heywood2016}. The tail of steep negative spectral indices visible in panel $d$ of Figure \ref{fig:stats} is almost certainly an artefact of these effects. We thus restrict interpretation of the spectral index measurements to the region where their distribution remains roughly symmetric at log$_{10}$ $S_{\mathrm{peak}}$~$>$~1.5. 

Pulsars are known to typically exhibit steep spectra \citep[$\alpha_{\mathrm{mean}}$~=~$-$1.6][]{jankowski2018}, and therefore may appear as outliers from the general population of compact radio sources, which is dominated by extragalactic synchrotron sources with a typical spectral index of $-$0.7. The error in the spectral index fit $\sigma_{\alpha}$, i.e.~a measure of how well the spectrum is captured by a simple power law, is shown in panel $e$.

\subsection{The MSP population and other outlier sources}

The temporal and spectral properties of the continuum counterparts of the known MSP population are plotted using blue markers on Figure \ref{fig:stats}. Deviations from the control sample are seen most readily in panels $a$ ($V$), $b$ ($\eta$), and $e$ ($\sigma_{\alpha}$). A linear fit to the distribution of these three parameters is shown by the solid line. The dashed lines are $\pm$2$\overline{\sigma}$, where $\overline{\sigma}$ is the mean of the standard deviations of the distribution computed in eight logarithmically-spaced peak brightness bins. These lines crudely delineates regions of the parameter space containing outliers. Plots of both the MSP and control sample lightcurves can be found online\footnote{\url{https://github.com/IanHeywood/globular}}.

Temporal variability in the MSP continuum emission could be explained by either differing diffractive interstellar scintillation conditions along the sightlines to the pulsars, intrinsic variability of the pulsars themselves, modulations due to binary orbits, or some combination of these. Approximately 20 per cent of the known MSPs could be identified as outliers in the $V$ and $\eta$ plots. The strong variability of MSPs C and J that was predicted via their associated imaging artefacts is confirmed, with these two sources being the most apparent. The distribution of the $\xi_{\mathrm{max}}$ statistic shows that at the 15 minute imaging cadence none of the MSPs appear as outliers, and thus do not have single epochs that deviate significantly from their general behavior. 

Investigation of the six components with log$_{\mathrm{10}}$~$\xi_{\mathrm{max}}$~$>$~1 in panel $c$ of Figure \ref{fig:stats} revealed that they are all pairs of sources that are close in projection one another. Scans that are at lower elevations exhibit larger, more elongated synthesised beams, and in those instances the emission from the sources encroaches on the position of the fainter one, causing the high $\xi_{\mathrm{max}}$ value when the photometry of the fainter source is extracted. Such instances could be automatically identified by considering the angular resolution of each individual image and the angular separations between a source and its neighbours, but we made no attempt to implement such a scheme here. The outlier source visible to the lower right of 47 Tuc C in the $V$ plot is also the partially blended source with the highest $\xi_{\mathrm{max}}$ value.

A far more effective way to identify the outlying MSPs is to examine the $\sigma_{\alpha}$ distribution, in which approximately half of the MSPs can be distinguished. This suggests that scintillation is the dominant source of the variability, as the interference pattern induced by the ionized material along the lines of sight to each pulsar introduces random fluctuations into the radio emission in frequency as well as time. This can also explain the spectral index values that are extremely steep (e.g. 47 Tuc H) and inverted (e.g. 47 Tuc J), as seen in panel $d$.

\subsection{J002402.7$-$720539.4: A candidate millisecond pulsar}
\label{sec:newpulsar}

% \begin{figure}
% \centering
% \includegraphics[width= 0.78 \columnwidth]{fig_candidate_lc.eps}
% \caption{Light curve for the candidate pulsar.}
% \label{fig:candidate}
% \end{figure}

The source from the control sample that is highlighted by the pink box in Figure \ref{fig:stats} occupies the same regions of parameter space as the known MSP population. We identify this as a candidate new MSP, or perhaps the localised radio continuum counterpart of one of the six MSPs in 47 Tuc that do not yet have good positions available from the timing solutions. These are P and V \citep{ridolfi2016}, ac and ad \citep{ridolfi2021}, and ae and af\footnote{\url{http://trapum.org/discoveries/}}. The radio position of the candidate is 1$''$ offset from the position of the X-ray source W52 \citep{grindlay2001}, which has an intrinsic X-ray luminosity in the 0.5--2.5 keV of 2.5~$\times$~10$^{30}$~erg~s$^{-1}$, consistent with the X-ray luminosities of the general MSP population in 47 Tuc.

\section{Summary}

We have investigated the image domain properties of the MSP population in 47 Tuc by exploiting the time and frequency dimensions of an archival MeerKAT observation. Approximately 20 per cent of the MSPs could be discovered using the temporal variability statistics that are generally employed in image domain searches for transients and variables. This fraction increases to approximately 50 per cent when the spectral domain is considered. We have identified one source as a candidate previously unknown MSP, or otherwise provide an interferometric localisation of one of the known MSPs in 47 Tuc for which no position is yet available through timing solutions. Scintillation is the likely cause of the MSPs appearing as outliers in the various metrics we have explored, with differing sightlines to the MSPs in the cluster resulting in a range of differing light curves. Modulation of the brightness due to binary orbits is hinted at in the light curve of at least one MSP (47 Tuc O), which is known to have a 3.12 hour orbital period that is coincident with peaks in the lightcurve \citep{freire2017}. Additional monitoring (possibly using MeerKAT's S-band to lessen the effects of scintillation) required to investigate this. S-band imaging would also potentially allow some of the blended pairs of MSPs to be resolved separately. 

The approach used in this paper is computationally cheaper than the true variance imaging method discussed by \citet{dai2016}, but is demonstrated to be a fruitful method for the detection of MSPs and other scintillating sources. We have made no attempt to optimise the time/frequency imaging intervals to match the scintillation scales in those domains. This may result in higher detection fractions, and will be investigated in a future work, which we will extend to include imaging of additional GCs and full Stokes imaging where possible. This work demonstrates that commensal imaging of fields that are being targeted for beamformer pulsar searches offers a potentially valuable way to localise pulsars immediately, as well as the potential for discovering new pulsars in image domain programs that fully exploit the time and frequency dimensions of the data.

\section*{Acknowledgements}

The MeerKAT telescope is operated by the South African Radio Astronomy Observatory, which is a facility of the National Research Foundation, an agency of the Department of Science and Innovation. This research made use of Astropy,\footnote{\url{http://www.astropy.org}} a community-developed core Python package for Astronomy \citep{astropy:2013, astropy:2018}. This work has made use of the Cube Analysis and Rendering Tool for Astronomy \citep[CARTA;][]{comrie2021}. This research has made use of NASA's Astrophysics Data System. I acknowledge support from the Breakthrough Listen initiative, the UK Science and Technology Facilities Council, and from the South African Radio Astronomy Observatory. Breakthrough Listen is managed by the Breakthrough Initiatives, sponsored by the Breakthrough Prize Foundation. I thank my colleagues in the Oxford pulsar and relativistic accretion groups for useful discussions. I thank Paulo Freire for maintaining a comprehensive web resource on pulsars in globular clusters which was useful for this work. Finally, I thank the anonymous referee for their useful comments.

%%%%%%%%%%%%%%%%%%%%%%%%%%%%%%%%%%%%%%%%%%%%%%%%%%
\section*{Data Availability}

The visibility data are available from the SARAO archive under the capture block ID listed in Table \ref{tab:meerkat_obs}. Tabular data and light curve images are available at \url{ https://github.com/IanHeywood/globular}. Synthesised images are available from the author.

%%%%%%%%%%%%%%%%%%%% REFERENCES %%%%%%%%%%%%%%%%%%

% The best way to enter references is to use BibTeX:

% Alternatively you could enter them by hand, like this:
% This method is tedious and prone to error if you have lots of references
%\begin{thebibliography}{99}
%\bibitem[\protect\citeauthoryear{Author}{2012}]{Author2012}
%Author A.~N., 2013, Journal of Improbable Astronomy, 1, 1
%\bibitem[\protect\citeauthoryear{Others}{2013}]{Others2013}
%Others S., 2012, Journal of Interesting Stuff, 17, 198
%\end{thebibliography}

%%%%%%%%%%%%%%%%%%%%%%%%%%%%%%%%%%%%%%%%%%%%%%%%%%

%%%%%%%%%%%%%%%%% APPENDICES %%%%%%%%%%%%%%%%%%%%%

% \appendix

% \section{1.28 GHz light curves for the MSPs in 47 Tuc and \omcen}

%%%%%%%%%%%%%%%%%%%%%%%%%%%%%%%%%%%%%%%%%%%%%%%%%%

% Don't change these lines
\bsp    % typesetting comment
\label{lastpage}

\begin{thebibliography}{}

% \bibliographystyle{mnras}
% \bibliography{example} % if your bibtex file is called example.bib

\bibitem[\protect\citeauthoryear{Abbate et al.}{2023}]{abbate2023} Abbate F., Possenti A., Ridolfi A., Venkatraman Krishnan V., Buchner S., Barr E.~D., Bailes M., et al., 2023, MNRAS, 518, 1642. doi:10.1093/mnras/stac3248

\bibitem[\protect\citeauthoryear{Andersson et al.}{2023}]{andersson2023} Andersson A., Lintott C., Fender R., Bright J., Carotenuto F., Driessen L., Espinasse M., et al., 2023, MNRAS.tmp. doi:10.1093/mnras/stad1298

\bibitem[\protect\citeauthoryear{Astropy Collaboration et al.}{2013}]{astropy:2013} Astropy Collaboration, Robitaille T.~P., Tollerud E.~J., et al., 2013, A\&A, 558, A33. doi:10.1051/0004-6361/201322068

\bibitem[\protect\citeauthoryear{Astropy Collaboration et al.}{2018}]{astropy:2018} Astropy Collaboration, Price-Whelan A.~M., Sip{\H{o}}cz B.~M., et al., 2018, AJ, 156, 123. doi:10.3847/1538-3881/aabc4f

\bibitem[\protect\citeauthoryear{Backer et al.}{1982}]{backer1982} Backer D.~C., Kulkarni S.~R., Heiles C., Davis M.~M., Goss W.~M., 1982, Natur, 300, 615. doi:10.1038/300615a0

\bibitem[\protect\citeauthoryear{Baumgardt \& Vasiliev}{2021}]{baumgardt2021} Baumgardt H., Vasiliev E., 2021, MNRAS, 505, 5957. doi:10.1093/mnras/stab1474

\bibitem[\protect\citeauthoryear{CASA Team et al.}{2022}]{casa2022} CASA Team, Bean B., Bhatnagar S., Castro S., Donovan Meyer J., Emonts B., Garcia E., et al., 2022, PASP, 134, 114501. doi:10.1088/1538-3873/ac9642

% \bibitem[\protect\citeauthoryear{Chen et al.}{2018}]{chen2018} Chen S., Richer H., Caiazzo I., Heyl J., 2018, ApJ, 867, 132. doi:10.3847/1538-4357/aae089

\bibitem[\protect\citeauthoryear{Caleb et al.}{2022}]{caleb2022} Caleb M., Heywood I., Rajwade K., Malenta M., Stappers B.~W., Barr E., Chen W., et al., 2022, NatAs, 6, 828. doi:10.1038/s41550-022-01688-x

\bibitem[\protect\citeauthoryear{Camilo et al.}{2000}]{camilo2000} Camilo F., Lorimer D.~R., Freire P., Lyne A.~G., Manchester R.~N., 2000, ApJ, 535, 975. doi:10.1086/308859

\bibitem[\protect\citeauthoryear{Chen et al.}{2023}]{chen2023} Chen W., Freire P.~C.~C., Ridolfi A., Barr E.~D., Stappers B., Kramer M., Possenti A., et al., 2023, MNRAS, 520, 3847. doi:10.1093/mnras/stad029

\bibitem[\protect\citeauthoryear{Comrie et al.}{2021}]{comrie2021} Comrie A., Wang K.-S., Hsu S.-C., et al., 2021, zndo, doi:10.5281/zenodo.3377984

% \bibitem[\protect\citeauthoryear{Dai et al.}{2020}]{dai2020} Dai S., Johnston S., Kerr M., Camilo F., Cameron A., Toomey L., Kumamoto H., 2020, ApJL, 888, L18. doi:10.3847/2041-8213/ab621a

% \bibitem[\protect\citeauthoryear{Dai et al.}{2023}]{dai2023} Dai S., Johnston S., Kerr M., Berteaud J., Bhattacharyya B., Camilo F., Keane E., 2023, MNRAS, 521, 2616. doi:10.1093/mnras/stad704

\bibitem[\protect\citeauthoryear{Dai et al.}{2016}]{dai2016} Dai S., Johnston S., Bell M.~E., Coles W.~A., Hobbs G., Ekers R.~D., Lenc E., 2016, MNRAS, 462, 3115. doi:10.1093/mnras/stw1871

\bibitem[\protect\citeauthoryear{Driessen et al.}{2020}]{driessen2020} Driessen L.~N., McDonald I., Buckley D.~A.~H., Caleb M., Kotze E.~J., Potter S.~B., Rajwade K.~M., et al., 2020, MNRAS, 491, 560. doi:10.1093/mnras/stz3027

\bibitem[\protect\citeauthoryear{Driessen et al.}{2022}]{driessen2022} Driessen L.~N., Stappers B.~W., Tremou E., Fender R.~P., Woudt P.~A., Armstrong R., Bloemen S., et al., 2022, MNRAS, 512, 5037. doi:10.1093/mnras/stac756

\bibitem[\protect\citeauthoryear{Grindlay et al.}{2001}]{grindlay2001} Grindlay J.~E., Heinke C., Edmonds P.~D., Murray S.~S., 2001, Sci, 292, 2290. doi:10.1126/science.1061135

\bibitem[\protect\citeauthoryear{Fender et al.}{2016}]{fender2016} Fender R., Woudt P.~A., Corbel S., Coriat M., Daigne F., Falcke H., Girard J., et al., 2016, mks..conf, 13. doi:10.22323/1.277.0013

\bibitem[\protect\citeauthoryear{Freire et al.}{2001}]{freire2001} Freire P.~C., Camilo F., Lorimer D.~R., Lyne A.~G., Manchester R.~N., D'Amico N., 2001, MNRAS, 326, 901. doi:10.1046/j.1365-8711.2001.04493.x

\bibitem[\protect\citeauthoryear{Freire et al.}{2017}]{freire2017} Freire P.~C.~C., Ridolfi A., Kramer M., Jordan C., Manchester R.~N., Torne P., Sarkissian J., et al., 2017, MNRAS, 471, 857. doi:10.1093/mnras/stx1533

\bibitem[\protect\citeauthoryear{Harris}{1996}]{harris1996} Harris W.~E., 1996, AJ, 112, 1487. doi:10.1086/118116

\bibitem[\protect\citeauthoryear{Heywood et al.}{2016}]{heywood2016} Heywood I., Jarvis M.~J., Baker A.~J., Bannister K.~W., Carvalho C.~S., Hardcastle M., Hilton M., et al., 2016, MNRAS, 460, 4433. doi:10.1093/mnras/stw1250

\bibitem[\protect\citeauthoryear{Heywood}{2020}]{heywood2020} Heywood I., 2020, ascl.soft. ascl:2009.003

\bibitem[\protect\citeauthoryear{Heywood et al.}{2022}]{heywood2022} Heywood I., Jarvis M.~J., Hale C.~L., Whittam I.~H., Bester H.~L., Hugo B., Kenyon J.~S., et al., 2022, MNRAS, 509, 2150. doi:10.1093/mnras/stab3021

\bibitem[\protect\citeauthoryear{Hotan et al.}{2021}]{hotan2021} Hotan A.~W., Bunton J.~D., Chippendale A.~P., Whiting M., Tuthill J., Moss V.~A., McConnell D., et al., 2021, PASA, 38, e009. doi:10.1017/pasa.2021.1

\bibitem[\protect\citeauthoryear{Howell, Guhathakurta, \& Gilliland}{2000}]{howell2000} Howell J.~H., Guhathakurta P., Gilliland R.~L., 2000, PASP, 112, 1200. doi:10.1086/316621

\bibitem[\protect\citeauthoryear{Hugo et al.}{2022}]{hugo2022} Hugo B.~V., Perkins S., Merry B., Mauch T., Smirnov O.~M., 2022, ASPC, 532, 541. doi:10.48550/arXiv.2206.09179

\bibitem[\protect\citeauthoryear{Hurley-Walker et al.}{2022}]{hurleywalker2022} Hurley-Walker N., Zhang X., Bahramian A., McSweeney S.~J., O'Doherty T.~N., Hancock P.~J., Morgan J.~S., et al., 2022, Natur, 601, 526. doi:10.1038/s41586-021-04272-x

\bibitem[\protect\citeauthoryear{Jankowski et al.}{2018}]{jankowski2018} Jankowski F., van Straten W., Keane E.~F., Bailes M., Barr E.~D., Johnston S., Kerr M., 2018, MNRAS, 473, 4436. doi:10.1093/mnras/stx2476

\bibitem[\protect\citeauthoryear{Jonas \& MeerKAT Team}{2016}]{jonas2016} Jonas J., MeerKAT Team, 2016, in \emph{MeerKAT Science: On the pathway to the SKA}, mks..conf, PoS, 1

\bibitem[\protect\citeauthoryear{Kenyon et al.}{2018}]{kenyon2018} Kenyon J.~S., Smirnov O.~M., Grobler T.~L., Perkins S.~J., 2018, MNRAS, 478, 2399. doi:10.1093/mnras/sty1221

\bibitem[\protect\citeauthoryear{Kurtzer, Sochat, \& Bauer}{2017}]{kurtzer2017} Kurtzer G.~M., Sochat V., Bauer M.~W., 2017, PLoSO, 12, e0177459. doi:10.1371/journal.pone.0177459

\bibitem[\protect\citeauthoryear{Manchester et al.}{1991}]{manchester1991} Manchester R.~N., Lyne A.~G., Robinson C., D'Amico N., Bailes M., Lim J., 1991, Natur, 352, 219. doi:10.1038/352219a0

\bibitem[\protect\citeauthoryear{Mohan \& Rafferty}{2015}]{mohan2015} Mohan N., Rafferty D., 2015, ascl.soft. ascl:1502.007

\bibitem[\protect\citeauthoryear{Murphy et al.}{2021}]{murphy2021} Murphy T., Kaplan D.~L., Stewart A.~J., O'Brien A., Lenc E., Pintaldi S., Pritchard J., et al., 2021, PASA, 38, e054. doi:10.1017/pasa.2021.44

\bibitem[\protect\citeauthoryear{Offringa et al.}{2014}]{offringa2014} Offringa A.~R., McKinley B., Hurley-Walker N., Briggs F.~H., Wayth R.~B., Kaplan D.~L., Bell M.~E., et al., 2014, MNRAS, 444, 606. doi:10.1093/mnras/stu1368

\bibitem[\protect\citeauthoryear{Oosterloo et al.}{2020}]{oosterloo2020} Oosterloo T.~A., Vedantham H.~K., Kutkin A.~M., Adams E.~A.~K., Adebahr B., Coolen A.~H.~W.~M., Damstra S., et al., 2020, A\&A, 641, L4. doi:10.1051/0004-6361/202038378

\bibitem[\protect\citeauthoryear{Pan et al.}{2016}]{pan2016} Pan Z., Hobbs G., Li D., Ridolfi A., Wang P., Freire P., 2016, MNRAS, 459, L26. doi:10.1093/mnrasl/slw037

\bibitem[\protect\citeauthoryear{Ransom}{2008}]{ransom2008} Ransom S.~M., 2008, IAUS, 246, 291. doi:10.1017/S1743921308015810

\bibitem[\protect\citeauthoryear{Ridolfi et al.}{2016}]{ridolfi2016} Ridolfi A., Freire P.~C.~C., Torne P., Heinke C.~O., van den Berg M., Jordan C., Kramer M., et al., 2016, MNRAS, 462, 2918. doi:10.1093/mnras/stw1850

\bibitem[\protect\citeauthoryear{Ridolfi et al.}{2021}]{ridolfi2021} Ridolfi A., Gautam T., Freire P.~C.~C., Ransom S.~M., Buchner S.~J., Possenti A., Venkatraman Krishnan V., et al., 2021, MNRAS, 504, 1407. doi:10.1093/mnras/stab790

\bibitem[\protect\citeauthoryear{Robinson et al.}{1995}]{robinson1995} Robinson C., Lyne A.~G., Manchester R.~N., Bailes M., D'Amico N., Johnston S., 1995, MNRAS, 274, 547. doi:10.1093/mnras/274.2.547

\bibitem[\protect\citeauthoryear{Rowlinson et al.}{2022}]{rowlinson2022} Rowlinson A., Meijn J., Bright J., van der Horst A.~J., Chastain S., Fijma S., Fender R., et al., 2022, MNRAS, 517, 2894. doi:10.1093/mnras/stac2460

\bibitem[\protect\citeauthoryear{Smirnov}{2011a}]{smirnov2011a} Smirnov O.~M., 2011, A\&A, 527, A107. doi:10.1051/0004-6361/201116434

\bibitem[\protect\citeauthoryear{Smirnov}{2011b}]{smirnov2011b} Smirnov O.~M., 2011, A\&A, 527, A108. doi:10.1051/0004-6361/201116435

\bibitem[\protect\citeauthoryear{Smirnov \& Tasse}{2015}]{smirnov2015} Smirnov O.~M., Tasse C., 2015, MNRAS, 449, 2668. doi:10.1093/mnras/stv418

\bibitem[\protect\citeauthoryear{Smirnov et al.}{2022}]{smirnov2022} Smirnov O.~M., Heywood I., Perkins S.~J., van Rooyen R., 2022, ASPC, 532, 385

\bibitem[\protect\citeauthoryear{Tasse et al.}{2018}]{tasse2018} Tasse C., Hugo B., Mirmont M., Smirnov O., Atemkeng M., Bester L., Hardcastle M.~J., et al., 2018, A\&A, 611, A87. doi:10.1051/0004-6361/201731474





\end{thebibliography}
\end{document}